\documentstyle[11pt]{article}

\begin{document}
\newcommand{\ECM}{\em Departament d'Estructura i Constituents de la
Mat\`eria \\
and\\
I. F. A. E.\\
Facultat de F\'\i sica, Universitat de Barcelona \\
Diagonal 647, E-08028 Barcelona, Spain}

\def\thefootnote{\fnsymbol{footnote}}
\pagestyle{empty}
{\hfill \parbox{6cm}{\begin{center} quant-ph/9803xxx\\
                                    March 1998\\
                     \end{center}}}
\vspace{1.5cm}

\centerline{
\large{Minimal optimal generalized quantum measurements}}

\vskip .6truein
\centerline {
J.I. Latorre\footnote{e-mail: latorre@ecm.ub.es},
P. Pascual\footnote{e-mail: pascual@ecm.ub.es}
and R. Tarrach\footnote{e-mail: tarrach@ecm.ub.es}}

\vspace{.3cm}
\begin{center}
\ECM
\\
\end{center}
\vspace{1.5cm}

\centerline{\bf Abstract}
\medskip

Optimal and finite positive operator valued measurements on a
finite number $N$ of identically  prepared systems
have been presented recently. With physical realization in mind we
propose here optimal and {\it minimal} generalized
quantum measurements for two-level systems.
 We explicitly construct them up 
to $N=7$
and verify that they are minimal up to $N=5$. We finally propose an
expression which gives the size of the minimal optimal measurements for
arbitrary $N$.

\newpage                                                                
\pagestyle{plain}

Consider a spin 1/2 particle (or any other two-level system)
which is in a pure state $\vert \Psi\rangle$ about which we do not know
anything,
that is, its spin points with equal probability into any direction. By
performing a measurement on the system one learns something about
$\vert\Psi\rangle$, that is, the $\it{a}$ $\it{priori}$ uniform probability distribution
becomes $\it{a}$ $\it{posteriori}$ a nonuniform distribution. Suppose now we have
$N$ identical copies of $\vert \Psi\rangle$, $\vert \Psi\rangle^{N} \equiv \vert
\Psi\rangle \otimes \vert \Psi\rangle \otimes \vert \Psi\rangle ... \otimes \vert \Psi\rangle$
($N$ times). Measurements on this enlarged system allow to learn more
about $\vert \Psi\rangle$. The amount of knowledge measurements allow to
extract from $\vert \Psi\rangle^{N}$ about $\vert \Psi\rangle$ is a monotonically
increasing function of $N$. Only in the limit $N \rightarrow \infty$ can
$\vert \Psi\rangle$ be determined exactly. This is because only in this limit are 
$\vert \Psi\rangle^{N}$ and $\vert \Psi'\rangle^{N}$ orthogonal whenever $\vert \Psi\rangle
\ne \vert \Psi'\rangle$, and thus distinguishable by an adequate
measurement.

For finite $N$ Massar and Popescu $\cite {MP}$ (see also
Holevo $\cite {H})$ obtained the $\it {optimal}$ measurement procedure
for spin 1/2 particles.
Their procedure, leading to the maximal knowledge about $\vert
\Psi\rangle$, corresponds to a positive operator valued measurement (POVM)
consisting of an $\it {infinite}$ isotropic set of projectors in the
Hilbert space of $\vert \Psi\rangle^{N}$. It is a measurement on the $\it
{combined}$ system. By Neumark's theorem $\cite {N}, \cite {P}$ this
corresponds to a von Neumann measurement in an infinitely dimensional
extension of the Hilbert space of $\vert \Psi\rangle^{N}$. This makes the
procedure academic, since it cannot be realized physically.\par

The next step was taken by Derka, Buzek and Ekert $\cite {DBE}$.
They explicitly construct an optimal $\it {finite}$ POVM, thus making the
procedure in principle accessible to the laboratory, and thus of relevance to
quantum computation and quantum communication. They quantify the acquired
knowledge about $\vert \Psi\rangle$ by the mean fidelity, $\overline {f}$,
whose
maximal value obtained by their procedure is                  
                                                                        
\begin{equation}                                                                     
\overline {f}_{max}={{N+1}\over {N+2}}.
\label{E01}
\end{equation}
Their POVM requires a finite number $n=(N+1)^2$ of projectors in the
Hilbert space of $\vert \Psi\rangle^{N}$. It is thus an optimal, finite,
generalized quantum measurement. But it is not minimal:      optimal
POVMs with a smaller number of projectors exist, as we will show. They
allow to learn the same by reading a smaller output. When it comes to
physical realizations this should be an advantage.\par

Here we present explicit results on optimal, finite and furthermore 
 minimal POVMs. The number of projectors $n$ they require is
roughly one third the number needed by the only optimal and finite
measurements known up to now $\cite {DBE}$. 
We have proceeded from N=2 up to N=5 case by case,
because we do not know how to build the POVM
algorithmically. They are optimal and
minimal. Then we construct optimal POVMs for $N=6$ and $N=7$ which we
strongly believe to be minimal. This belief is 
based on a bit of mathematical intuition and some numerical
frustration, but we have not been able to rigorously
exclude POVMs with one projector less. We finally propose and explain
a formula which gives the minimal $n$ as a function of $N$ and 
which reproduces all our explicit results. 

\bigskip

Let us first introduce some notation (we will try to follow
reference \cite{DBE} whenever possible). Our POVM is given by a
 finite set of $n$ one
dimensional projectors built from the states of maximal spin, $s={N\over
2}$, and maximal spin component in some direction,

\begin{equation}
\vert\theta_r, \psi_r\rangle^N\equiv \vert\theta_r,\psi_r\rangle
\otimes\vert\theta_r,\psi_r\rangle
\otimes...\otimes\vert\theta_r,\psi_r\rangle, \qquad r=1,...n, 
\label{E3}
\end{equation}
where $\vec {\sigma}\cdot\hat n(r) \vert\theta_r,\psi_r\rangle 
= \vert\theta_r,\psi_r\rangle,
\hat n(r) =(\sin \theta_r \cos \psi_r, \sin \theta_r \sin \psi_r, \cos
\theta_r)$ and such that

\begin{equation}
\sum^n_{i=1}\ c^2_r\ \vert\theta_r,
\psi_r\rangle^{N}{}^{N}\langle\theta_r,\psi_r\vert= I^{(s={N\over 2})},\qquad 0<c^2_r\leq 1.
\label{E4}
\end{equation}
Here the r.h.s. represents the identity in the maximal spin space.
Notice that $n$ has to be larger than the dimension of the maximal spin
space, $N+1$, as $n=N+1$ would require the $n$ projectors of Eq.(\ref{E4}) to
be orthogonal, which they are not. The extension of Eq.(\ref{E4}) to the
complete $2^N$-dimensional Hilbert space is straightforward, but
irrelevant, as the corresponding projectors, being orthogonal to
$\vert\Psi\rangle^N$, do not allow to increase our knowledge about
$\vert\Psi\rangle$.

We know from references \cite{MP}, \cite{H}
 and \cite{DBE} that a POVM of the type we
are considering is optimal. This means that the mean fidelity,

\begin{equation}
\overline {f}\equiv\sum^n_{r=1}\int D \hat
{n} \ \ \vert {}^N\langle\Psi\vert\theta_r,\psi_r\rangle^N
\vert^2
\ c^2_r\ \vert\langle\Psi\vert\theta_r,\psi_r\rangle\vert^2,
\label {E5}
\end{equation}
where $\vert\Psi\rangle\equiv \vert\theta, \psi\rangle=\vec {\sigma}\cdot \hat {n}
\vert\theta, \psi\rangle, \hat {n}=(\sin\theta \cos\psi, \sin\theta 
\sin\psi,
\cos\theta)$ and the isotropic measure is such that

\begin{equation}
\int D\hat {n}\ \ \vert\theta, \psi\rangle\langle\theta, \psi\vert=
{1\over 2} \pmatrix {1 & 0 \cr
0 & 1\cr},\ 
\label {E6}
\end{equation}
is maximal, see Eq.(\ref{E01}). It was also shown in reference 
\cite{DBE} that for
optimal POVMs Eq.(\ref{E4}) can be substituted by the much simpler one

\begin{equation}
\sum^n_{r=1}\ c^2_r\ \ \vert {}^N\langle\theta,
 \psi\vert\theta_r,\psi_r\rangle^N\vert^2=1,
\qquad\forall\vert\theta, \psi\rangle.
\label {E7}
\end{equation}
This is therefore the equation we want to study and solve, i.e. find
$c^2_r$, $\theta_r$ and $\psi_r$, $r=1, 2...n$, for the
smallest $n$ possible.

It is not difficult to prove from the explicit expression for
$\vert {}^N\langle\theta, \psi\vert\theta_r,\psi_r
\rangle^N\vert^2$ and expanding monomials
in terms of Legendre polynomials that Eq. (\ref{E7}) is equivalent to
\begin{eqnarray}
\sum^n_{r=1}\ c^2_r &=&N+1 \nonumber \\ 
\sum^n_{r=1}\ c^2_r \ P^M_L (\cos\theta_r)e^{iM\psi_r}&=&0,
\qquad L=1,...N,\qquad M=0,...L,
\label{E8}
\end{eqnarray}
where the dependence on $\theta$, $\psi$ has been traded for a set of
equations. Again, after some algebra, this set of equations can be shown
to be equivalent to
\begin{eqnarray}
\sum^n_{r=1}\ c^2_r \ z^k_r \ x^m_r&=& {1+(-1)^k\over 2}{1+(-1)^m\over 2}
(N+1) {{(m-1)!!(k-1)!!}\over (k+m+1)!!} \nonumber \\ 
\sum^n_{r=1}c^2_r z^k_r x^{m-1}_r y_r&=&0,\qquad m\geq 1,
\label{E9}
\end{eqnarray}
where $m=0,... N$, $k=0,... N-m$, $(-1)!!=1$ and $\hat n(r)\equiv (x_r,
y_r, z_r)$. Finally, another equivalent set of equations, which we
have found most useful, is
\begin{eqnarray}
&&\sum^n_{r=1} \ c^2_r \ = N+1 \nonumber\\
&&\sum^n_{r=1} \ c^2_r \  n_\alpha(r)=0 \nonumber\\
&&\sum^n_{r=1} \ c^2_r \  n_\alpha(r) n_\beta(r) = {N+1\over 3}
\delta_{\alpha\beta} \nonumber\\
&&\sum^n_{r=1} \ c^2_r \ n_\alpha(r) n_\beta(r) n_\gamma(r)=0 \nonumber\\
&&\dots
\label{E99}
\end{eqnarray}
which in compact form reads
\begin{equation}
\sum^n_{r=1} c^2_r \hat n(r)^q= {1+(-1)^q\over 2} {N+1\over q+1} I^{(q)},
\qquad q=0,... N,
\label{E10}
\end{equation}
where $\hat n(r)^q \equiv \hat n(r) \otimes \hat n(r) \otimes ...
\otimes \hat n(r)$ with $q$ factors, and $I^{(q)}$ is the invariant
symmetric rank $q$ tensor, trace-normalized to $q+1$,
$I^{(0)}\equiv 1$,  
$I^{(2)}_{\alpha\beta}\equiv
\delta_{\alpha\beta}$, 
$I^{(4)}_{\alpha\beta\gamma\delta}\equiv
{1\over 3}(\delta_{\alpha\beta}
 \delta _{\gamma\delta} +
\delta_{\alpha\gamma} \delta_{\beta\delta} + \delta_{\alpha\delta}
 \delta_{\beta\gamma})$, etc. 
In order to simplify our future
discussion we also note that Eq.(\ref{E10}) can be contracted 
with  $\hat n(i)^q$ leading to
\begin{equation}
\sum^n_{r\not= i}c^2_r {(\hat n(r) \cdot \hat n(i))}^q= {1+(-1)^q\over
2}{N+1\over q+1} - c^2_i, \qquad i=1,...n, \qquad q=0,...N.
\label{compacteq}
\end{equation}

\bigskip
Let us pause and reflect on the meaning of the above set of equations.
As $N$ increases, more equations in the hierarchy of 
Eq.(\ref{E99}) must be verified forcing that the distribution of $c_r^2$
 and
$\hat n(r)$  approach the form of a continuous uniform angular
distribution. Thus, for  finite $N$, we do expect to obtain highly 
symmetric solutions. 
 No algorithm to find out the minimal $n$ which produces 
a solution to the truncated set of equations has emerged from our
efforts. We
have, therefore, proceeded case by case from N=2 upwards.

\bigskip

Let us discuss in some detail
 the deduction of the explicit solution in the case $N=2$.
We have to solve the  first three set of equations in Eq.(\ref{E99}) for
the minimal  possible $n$. Using Eq.(\ref{compacteq})
the manifestly non-negative combination

\begin{equation}
S \equiv \sum^n_{r\not= i}\ c^2_r \ (b_i+\hat n(i) \cdot \hat n(r))^2
= b_i^2 (3-c^2_i) - 2 b_i c^2_i + 1 - c^2_i
\geq 0, \qquad
\forall i=1,...n.
\label{E14}
\end{equation}
can be evaluated. It reaches its minimum for 
\begin{equation}
b_i= {c^2_i\over 3 -c^2_i}
\end{equation}
giving
\begin{equation}
S= {3- 4 c^2_i\over 3 - c^2_i} \geq 0.
\end{equation}
This forces $c^2_i\leq 3/4$ and, furthermore,
\begin{equation}
\sum_{i=1}^n \left(3-4 c^2_i\right)= 3 (n-4) \geq 0,
\end{equation}
proving that $n\geq 4$. It is easy to see that a solution that
saturates the bound exists. Indeed, taking the largest possible value
for all $c^2_i$, that is $c^2_i=3/4$,  in our original
expression for $S$ we get
\begin{equation}
S= {3\over 4} \sum_{r\not= i}^n \left({1\over 3} +
\hat n(i) \cdot \hat n(r)\right)^2 =0,
\end{equation}
which implies that every term in the sum must vanish 
and leads to the final result
\begin{eqnarray}
&&n_{\it min}(N=2)=4\nonumber\\
&&c^2_i={3\over 4} \qquad i=1,\dots,4 \nonumber\\
&&\hat n(i) \cdot \hat n(j)=-{1\over 3} \qquad \forall i\not=j
\end{eqnarray}
This solution corresponds to a regular tetrahedron. The minimal
optimal POVM 
for N=2 is
thus organized as a Platonic polyhedron, $c^2_i$ playing the role of
the distance to the vertices from the center and $\hat n(i)$
pointing into the directions of the vertices. As anticipated, this
solution is unique by construction and stands as the smallest
 discretization
of angular integration.

\bigskip

The key idea to find out the above solution was to select a manifestly
positive combination of all the equations needed at level $N$. Let us
take advantage of this clue in  the case $N=3$, which corresponds
to solving the first four sets of equations in Eq. (\ref{E99}). We
combine them into the, again, manifestly non-negative expression
\begin{equation}
S\equiv \sum^n_{r\not= i}c^2_r (1+\hat n(i)\cdot \hat n(r)){(b_i+\hat n(i)\cdot
\hat n(r))}^2= 
b_i^2 (4-2 c_i^2)+2 b_i ({4\over 3}-2 c_i^2)+({4\over 3}-2 c_i^2)
 \geq 0, \qquad \forall i=1,...n.
\label {E17}
\end{equation}
The minimum of $S$ corresponds to
\begin{equation}
b_i=-{1\over 3}{2-3 c^2_i\over 2 - c^2_i}\qquad \Rightarrow \qquad
S={8\over 9}{2 -3 c^2_i\over 2 - c_i^2}\ .
\end{equation}
We, thus,  deduce that all $c_i^2\leq 2/3$, and 
\begin{equation}
\sum_{i=1}^n (2-3 c^2_i)=2 (n - 6) \geq 0.
\end{equation}
The bound is then $n\geq 6$. A solution that saturates the bound exists
and can be found by setting all $c^2_i=2/3$, leading to 
\begin{equation}
S=\sum_{r\not= i} \ c_r^2 (1+ \hat n(i)\cdot \hat n(r))
(\hat n(i)\cdot \hat n(r))^2=0\ .
\end{equation}
Every term in the sum must vanish; thus,
the scalar products of any pair of vectors are constrained to
\begin{equation}  
\hat n(i)\cdot \hat n(r)=\cases{0\cr -1\cr} 
\end{equation}
It is easy to use Eq.(\ref{E99})  to show that 
\begin{eqnarray}
&&n_{\it min}(N=3)=6\nonumber\\
&&c^2_i={2\over 3} \qquad i=1,\dots,6 \nonumber\\
&&\hat n(i) \cdot \hat n(j)=0
 \qquad \forall i\not=j\qquad {\rm except}\qquad 
\hat n(1) \cdot \hat n(6)=\hat n(2) \cdot \hat n(4)=
\hat n(3) \cdot \hat n(5)=-1\ .
\end{eqnarray}
This solution
corresponds to a regular octahedron. Once again a
Platonic polyhedron 
underlies the unique, optimal and minimal POVM for $N=3$.

\bigskip

At this point the reader may be wondering about the  role of
Platonic solids in constructing minimal POVMs. An immediate objection
arises from the fact that there are only a
finite number of Platonic solids, yet 
a vast series of more exotic, still highly symmetric, solids 
may take over as an organizing principle. 
It turns out that already at $N=4$ a more elaborated solution 
is found.

For $N=4$ we have found it convenient to start from
\begin{equation}
\sum^n_{r\not= i}c^2_r {(b_i + d_i \hat n(i)\cdot \hat n(r) +
{(\hat n(i)\cdot \hat n(r))}}^2)^2\geq 0\ .
\label {E20}
\end{equation}
Minimization with respect to $b_i$ and $d_i$ eventually lead to
\begin{equation}
\left({5\over 4}-c^2_i  \right) \left({5\over 9}-c^2_i  \right) \geq 0\ .
\label {E21}
\end{equation}
and
\begin{equation}
\sum_{i=1}^n \left({5\over 9}-c^2_i  \right)= 5 (n-9) \geq 0\ ,
\label {E211}
\end{equation}
which
implies $n\geq 9$. For $n=9$, 
the values obtained for $c^2_i$,
$c^2_i={5\over 9}$, and $\hat n_i\cdot \hat n_r$, from saturating
the bound, do not satisfy Eq.(\ref{E99}). Thus $n>9$ strictly.
Analyzing more elaborated
bounds, we have been able to prove that
for $n=10$ necessarily the $c_i^2$ cannot  all be identical.
By means of numerical inspiration, we have  found an explicit solution for 
$n=10$. Two of the $c_i^2$ 
turn out to be equal and smaller than the rest,
which are also equal among them,  and the $\hat n(i)$ point
to the vertices of a figure made as a twisted prism with pyramidal
caps (its explicit form is given later in the table).
We have therefore encountered a somewhat irregular but minimal 
solution to the POVM in the $N=4$ case. The {\it modus operandi}
is always related to exploiting a manifestly non-negative combination
of all the equations to be
solved.

\bigskip

For $N=5$ our starting point is
\begin{equation}
\sum^n_{r\not= i}c^2_r (1+\hat n_i\cdot \hat n_r) {(b_i + d_i
\hat n_i\cdot \hat n_r + {(\hat n_i\cdot \hat n_r)^2})^2}\geq 0\ ,
\label {E23}
\end{equation}
which after minimization leads to
\begin{equation}
\left(c^2_i - {1\over 2}\right)
 \geq 0\qquad
\Rightarrow \qquad \sum_{i=1}^n \left(1-2 c_i^2\right)=n-12\geq 0\ .
\label {E24}
\end{equation}
Thus $n\geq 12$. For $n=12$ we obtain a solution that does 
saturate the bound (in analogy to  $N=2,3$).
The explicit, 
unique, minimal solution is made with all $c^2_i=1/2$ and
$\hat n(i)\cdot\hat n(j)=-1,1/{\sqrt 5},-1/{\sqrt 5}$. Again, we defer
the detailed structure of the solution to the table.

\bigskip
Starting from expressions like Eqs. (\ref{E20}) and (\ref{E23}), but with a cubic
instead of quadratic polynomial, one can prove that $n > 16$ and
$n > 20$ for $N=6$ and $7$ respectively. Exhaustion has prevented us
from filling the gap between these lower bounds and the solutions with
$n=18$ and $n=22$ respectively,
which we have been able to build explicitly.
 Notice that of the four cases $N=2, 3,
4$ and $5$ for which we give a complete proof, for three of them, all
but $N=4$, our solution is also unique and corresponds to
constant $c^2_r$.

\begin{table}[h]
\begin{center}
\begin{tabular}{|c|c|c|c|c|}
\hline
&&&&\\
$N$&$n_{min}$&$c_r^2$&$\theta_r$&$\psi_r$\\
&&&&\\
\hline
$2$&
$4$&
$c_r^2= {3\over 4} \quad r=1..4$&
$\theta_1=0$&
$\psi_1=0$\\ 
&&&$\theta_r=\arccos({-1\over 3}) \quad r=2..4$&
$\psi_r=(r-2) {2\pi \over 3}\quad r=2,3,4$\\
\hline
$3$&
$6$&
$c_r^2= {2\over 3} \quad r=1..6$&
$\theta_1=0 \quad \theta_2=\pi$&
$\psi_1=0 \qquad \psi_2=0$\\ 
&&&$\theta_r={\pi\over 2} \quad r=3..6$&
$\psi_r=(r-3) {\pi \over 2}\quad r=3..6$\\
\hline
&
&
$c_1^2=c_2^2={5\over 12}$&
$\theta_1=0 \quad \theta_2=\pi$&
$\psi_1=0 \qquad \psi_2=0$\\ 
$4$&
$10$&
$c_r^2= {25\over 48} \quad r=3..10$&
$\theta_r=\arccos{1\over\sqrt{5}} \quad r=3..6$&
$\psi_r=(r-3) {\pi \over 2}\quad r=3..6$\\
&&&
$\theta_r=\pi-\theta_3 \quad r=7..10$&
$\psi_r=(r-{13\over 2}) {\pi \over 2}\quad r=7..10$\\
\hline
&
&
&
$\theta_1=0 \quad \theta_2=\pi$&
$\psi_1=0  \qquad \psi_2=0$\\ 
$5$&
$12$&
$c_r^2={1\over 2} \quad r=1..12$&
$\theta_r=\arccos{1\over\sqrt{5}} \quad r=3..7$&
$\psi_r=(r-3) {2\pi \over 5}\quad r=3..7$\\
&&&
$\theta_r=\pi-\theta_3 \quad r=8..12$&
$\psi_r=(r-{15\over 2}) {2\pi \over 5}\quad r=8..12$\\
\hline
\end{tabular}
\end{center}
\end{table}

\begin{table}[h]
\begin{center}
\begin{tabular}{|c|c|c|c|}
\hline
&
&
$\theta_1=0 \quad \theta_2=\pi$&
$\psi_1=0  \qquad \psi_2=0$\\ 
&
$c_1^2=c_2^2={14\over 45}$&
$\theta_r=\arccos{\sqrt{13+2\sqrt{30}}\over 7} \quad {}_{r=3..6}$&
$\psi_r=(r-3) {\pi \over 2}\quad {}_{r=3..6}$\\
$N=6$&
$c_r^2={7(410+\sqrt{30})\over 7200} \quad {}_{r=3..10}$&
$\theta_r=\pi-\theta_3 \quad {}_{r=7..10}$&
$\psi_r=(r-{13\over 2}) {\pi \over 2}\quad {}_{r=7..10}$\\
$n=18$&
$c_r^2={7(410-\sqrt{30})\over 7200} \quad {}_{r=11..18}$&
$\theta_r=\arccos{-\sqrt{13-2\sqrt{30}}\over 7}\quad {}_{r=11..14}$&
$\psi_r=\psi_{r-8}\quad {}_{r=11..18}$\\
&&
$\theta_r=\pi-\theta_{11}\quad {}_{r=15..18}$&\\
\hline
&
&
$\theta_1=0 \quad \theta_2=\pi$&
$\psi_1=0   \qquad \psi_2=0$\\ 
&
$c_1^2=c_2^2={10\over 27} $&
$\theta_r=\arccos{1\over 2}\sqrt{1+3\sqrt{3\over 35}} \quad {}_{r=3..7}$&
$\psi_r=(r-3) {2\pi \over 5}\quad {}_{r=3..7}$\\
$N=7$&
$c_r^2={147+\sqrt{105}\over 405} \quad {}_{r=3..12}$&
$\theta_r=\pi-\theta_3
 \quad r=8..12$&
$\psi_r=(r-{15\over 2}) {2\pi \over 5}\quad {}_{r=8..12}$\\
$n=22$&
$c_r^2={147-\sqrt{105}\over 405} \quad {}_{r=13..22}$&
$\theta_r=\arccos{-1\over 2}\sqrt{1-3\sqrt{3\over 35}} 
\quad {}_{r=13..17}$&
$\psi_r=\psi_{r-10}\quad {}_{r=13..22}$\\
&&
$\theta_r=\pi-\theta_{13} \quad {}_{r=18..22}$&
\\
\hline
\end{tabular}
\end{center}
\end{table}

We have summarized all our results in the above two tables.
We have also checked that they all satisfy the equations for
optimal POVMs of reference \cite{DBE}.
Having in our hands all these concrete solutions it is
possible to speculate on which  $n_{min}$ corresponds to a given $N$.
The formula we propose is
\begin{equation}
n_{min}(N)={\rm min} \left(1+ \left[{2+(N+1)^2\over 3}\right],
 4+2 \left[{N\over 2}\right]+ 2 \left[{2\over
3} \left[{N\over 2}\right]^2\right]\right)\ ,
\label {E2}
\end{equation}
where square brackets mean integer part.
To justify it,
let us first note that the
 number of independent equations in Eqs. (\ref{E8}), (\ref{E9}) or 
(\ref{E10}) is
$(N+1)^2$. 
The number of unknown variables in these equations is $3n-3$, where
rotational invariance has been used to fix $x_1=y_1=y_2=0$. Let us
clearly state that the problem of finding rigorously the minimal $n$
which for each $N$ allows to solve the non-linear system of 
Eq.(\ref{E99}) is
beyond our mathematical skills. 
However,  the explicit cases
 $N=2$ to 7 seem to suggest  that for this system one
can always find a solution when the number of unknown variables is at
least equal to the number of equations,
\begin{equation}
3n-3 \geq (N+1)^2
\label {E11}
\end{equation}
The minimal $n$ satisfying Eq. (\ref{E11}) leads to the first expression in Eq. (2).
On the other hand, limiting ourselves to solutions with even $n$ and for
which $\hat {n}_r + \hat n_{r-1}=0$, $c^2_r=c^2_{r-1}$, $r=2, 4 ... n$,
the system of Eq. (\ref{E99}) reduces then to its even $q$ part. The assumption
that the number of variables is at least the number of equations,

\begin{equation}
{3n\over 2} - 3 \geq 1 + 3 \left[{N\over 2}\right] +
 2  \left[{N\over 2}\right]^2
\label {E12}
\end{equation}
now leads to a minimal even $n$ given by the second expression in Eq. 
(\ref{E2}).
This is the justification of Eq. (\ref{E2}). It gives $n_{min}(6)=18$
and $n_{min}(7)=22$, which precisely corresponds to the minimal
solutions which we have been able to construct.

This means that one can do with roughly one third
the number of projectors required by the procedure of reference 
\cite{DBE}.
It turns out that for $N$ even the minimum is the first expression and
for $N$ odd the second. Also $n_{min}$ is always even.

Let us wind up with two comments. First, we have concentrated here on
optimal POVMs. We will come back, somewhere else, to optimal von Neumann
measurements. These are only known to exist for $N=2$ \cite{MP},
but we understand that the problem remains open for $N>2$.
Second, we have used here the mean fidelity as a measure of acquired
knowledge, but we could have used the more information-theoretic
decrease in Shannon entropy, as e.g. done in 
a related problem by Peres and
Wootters \cite{PW}. The maximal mean acquired knowledge, in bits, then
reads
\begin{equation}
\overline {\Delta I}= {1\over \ln 2}(\ln(N+1) - {N\over {N+1}})
\label {E13}
\end{equation}
Our conclusion would have been the same: we would have built the same
optimal, minimal, POVMs.

\vspace*{1cm}
\section{Acknowledgments}

Financial support from CICYT, contract AEN95-0590,
and from CIRIT, contract 1996GR00066 are
acknowledged.


\begin{thebibliography}{99}
\bibitem{MP} S. Massar and S. Popescu, {\sl Phys. Rev. Lett.}, {\bf 74},
1259 (1995).
\bibitem{H} A.S. Holevo, {\sl Probabilistic and Statistical Aspects of Quantum
Theory} (North-Holland, Amsterdam, 1982).
\bibitem{N} M.A. Neumark, {\sl C.R. Acad. Sci. URSS} {\bf 41}, 359
(1943).
\bibitem{P}A. Peres, {\sl Found. Physics} {\bf 20}, 1441 (1990).
\bibitem{DBE}R. Derka, V. Buzek and A.K. Ekert, {\sl Phys. Rev. Lett.}
{\bf 80}, 1571 (1998).
\bibitem{PW}A. Peres and W.K. Wootters, {\sl Phys. Rev. Lett.} {\bf 66},
1119 (1991).
  


\end{thebibliography}
\end{document}